\documentclass[review,authoryear,12pt]{elsarticle}

\usepackage{amssymb}

 \usepackage{multirow}
 
 \usepackage{caption}

\journal{Ultrasound in Medicine and Biology}

\begin{document}
\begin{frontmatter}

%% Title


\title{The Kidneys Are Not All Normal: Investigating the Speckle Distributions of Transplanted Kidneys}

\author[Affil1]{Rohit Singla \fnref{fn1} \corref{cor1}}
\author[Affil2]{Ricky Hu}
\author[Affil3]{Cailin Ringstrom}
\author[Affil3]{Victoria Lessoway}
\author[Affil3]{Janice Reid}
\author[Affil4]{Christopher Nguan}
\author[Affil3,Affil5]{Robert Rohling}

\address[Affil1]{School of Biomedical Engineering, University of British Columbia, Vancouver, Canada}
\address[Affil2]{Faculty of Medicine, Queen's University, Kingston, Canada}
\address[Affil3]{Electrical and Computer Engineering, University of British Columbia, Vancouver, Canada}
\address[Affil4]{Urologic Sciences, University of British Columbia, Vancouver, Canada}
\address[Affil5]{Mechanical Engineering, University of British Columbia, Vancouver, Canada}

\cortext[cor1]{Corresponding Author: Rohit Singla, 2332 Main Mall, V6T 1Z4; Email, rsingla@ece.ubc.ca}

\fntext[fn1]{R. Singla and R. Hu contributed equally to this work.}

\begin{abstract}
Modelling ultrasound speckle has generated considerable interest for its ability to characterize tissue properties. As speckle is dependent on the underlying tissue architecture, modelling it may aid in tasks like segmentation or disease detection. However, for the transplanted kidney where ultrasound is commonly used to investigate dysfunction, it is currently unknown which statistical distribution best characterises such speckle. This is especially true for the regions of the transplanted kidney: the cortex, the medulla and the central echogenic complex. Furthermore, it is unclear how these distributions vary by patient variables such as age, sex, body mass index, primary disease, or donor type. These traits may influence speckle modelling given their influence on kidney anatomy. We are the first to investigate these two aims. N=821 kidney transplant recipient B-mode images were automatically segmented into the cortex, medulla, and central echogenic complex using a neural network. Seven distinct probability distributions were fitted to each region. The Rayleigh and Nakagami distributions had model parameters that differed significantly between the three regions ($p \leq 0.05$). While both had excellent goodness of fit, the Nakagami had higher Kullbeck-Leibler divergence. Recipient age correlated weakly with scale in the cortex ($\Omega$: $\rho$ = 0.11, $p$ = 0.004), while body mass index correlated weakly with shape in the medulla (m: $\rho$ = 0.08, $p$ = 0.04). Neither sex, primary disease, nor donor type demonstrated any correlation. We propose the Nakagami distribution be used to characterize transplanted kidneys regionally independent of disease etiology and most patient characteristics based on our findings.
\end{abstract}

\begin{keyword}
ultrasound \sep speckle \sep kidney \sep B-mode \sep envelope statistics \sep quantitative ultrasound
\end{keyword}

\end{frontmatter}

%% Do not remove the page break here.
\pagebreak

%% MAIN TEXT INSTRUCTIONS

%% Commands for figures and tables should not be included in the main body of the submitted version of this file (e.g. the figure and tabular environments).  Figure captions should be listed in this file, as shown below.  Tables and Table captions should be listed as a separate section at the end of this file, as shown below.  Many authors may wish to include figures and tables within the main text of their document will preparing their manuscript.  This may be done, however please comment out any of the lines prior to submission.

%% Because the Elsevier editorial process does not allow for the figure and tabular environments in the submitted document, you will be unable to use autonumbering (i.e. \label and \ref) for figures and tables. 

%%  If long equations are used in the document, authors should use a two column format to make sure that the equations will break at approximately the right places.  To do this, replace the class option 'review' with the following two class options '3p' and 'twocolumn'.  Keep in mind that the column width produced in '3p' is slightly narrower than the final printer version.  After inserting the appropriate line breaks in your equation, change the '3p' option back to 'review'.

%% For citations, use the commands \citep and \citet

%% BEGIN MAIN TEXT

%%%%%%%%%%% INTRODUCTION
\section*{Introduction}
\label{intro}
At the microscopic level, inhomogeneities in tissue acoustic impedances result in scattered ultrasonic waves in ultrasound imaging.\citep{szabo2004diagnostic,burckhardt1978speckle} As the ultrasonic pulse is transmitted through tissue and interacts with it, the pulse is scattered in multiple directions. As a result of constructive and destructive interference, these scattered waves are visually depicted as “speckle” giving conventional B-mode ultrasound its characteristic grainy appearance. This speckle is both deterministic as well as non-random in nature.\citep{szabo2004diagnostic} Due to the fact that speckle is in part tissue-dependent, modelling speckle may allow for the subsequent characterization of tissue architecture or information related to regional morphology and signal scatterers.\citep{oelze2016review} The analysis of scattered waves may offer additional non-invasive information about the presence and type of disease in subsurface tissues in a way that conventional B-mode imaging does not.\citep{oelze2016review} This characterization requires either modelling the radiofrequency echoes or the compressed enveloped echoes.\citep{burckhardt1978speckle, destrempes2010critical, wagner1983statistics} While the radiofrequency echo data is from an earlier stage in image formation, availability is restricted to a few research-oriented ultrasound machines such as the Vantage system (Verasonics, Kirkland, WA). Contrarily, the B-mode image is the result of signal processing and filtering of the radiofrequency data, producing B-mode images. B-mode images are routinely used in clinical practice and stored in archiving systems; radiofrequency data is not. As a result, B-mode images is used in this work. 

Numerous models have been explored to characterize the envelope signal of ultrasound in the medical context. These commonly draw from communications theory and explore how to model a radiofrequency wave’s propagation given characteristics about its environment. \citeauthor{shankar1996studies} characterized breast ultrasound images of n=19 patients to discriminate between benign and malignant tumours.\citep{shankar1996studies} In the liver, \citeauthor{tsui2016acoustic} applied modelling of speckle for characterization of disease as compared to biopsy.\citep{tsui2016acoustic} They achieved an AUROC of 0.88 for the detection of fibrosis.\citep{tsui2016acoustic} Also in the liver, \citeauthor{wan2015effects} investigated 107 patients and discovered a significant positive correlation ($\rho$ = 0.84) between the organ's fat content and the Nakagami distribution parameters.  \citeauthor{nasr2018k} use B-mode images to differentiate between freshly excised and decellularized rodent kidneys.\citep{nasr2018k} Using a three-parameter model based on this data, they achieve an accuracy of 92\%.

These works serve to indicate that it is possible to quantify changes in tissue architecture and correlate them with speckle changes. In the human transplanted kidney, where ultrasound is the first line imaging investigation for kidney dysfunction \citep{taffel2017acr,singla2022kidney}, expanding the ability to non-invasively characterize the transplant kidney may be beneficial. Subtle changes in the kidney, whether it be distinguishing between regions or detecting disease, may be reflected in speckle. Such a characterization may assist in segmentation of the kidney and its regions, the detection of disease across a spectrum of etiologies, or provide prognostication for transplant organ outcomes. However, it remains to be seen which statistical distribution may best model the speckle.

As the kidney has a heterogeneous radial architecture, the speckle distributions within each of its regions may differ.\citep{netter2018atlas} The cortex, the outer region of the parenchyma which lies beneath the kidney’s capsule, contains glomeruli and convoluted tubules.\citep{netter2018atlas} On the other hand, the medulla, the inner region of the parenchyma, contains pyramidal structures and collecting tubules.\citep{netter2018atlas} The innermost compartment of the kidney, the central echogenic complex, is a combination of the sinus, fat, and vessels.\citep{netter2018atlas} Theoretically, these three regions present with different scattering properties given differences in their structure. For example, in response to an chronic injury such as hypertension, the kidney undergoes a mal-adaptive immune response. As part of this response and the healing process, glomeruli increase in density and size (glomerulosclerosis and hypertrophy). Changes in the glomeruli may be reflected in the speckle content of the cortex, such as how fat content influences the liver's speckle. This is supported by early work that suggests the kidney's glomeruli may be the predominant scatterers at clinical scanning frequencies. \citep{insana1995modeling,insana1991identifying,insana1992identifying}. The authors also suggest that vasculature, found at the interface between the cortex and medulla, may be a source of scattering at higher frequencies.

Furthermore, if speckle modelling is to be used widely in the transplanted kidney, it is important to consider the heterogeneity of the patient population. It is necessary to understand if and how the probability distributions differ across different recipient traits such as age, sex, body habitus, the primary diagnosis leading to kidney stage, as well as across donor traits such as type and age. These traits may influence the speckle given their influence on the anatomy and physiology of the kidney. 

This study explores which statistical distribution best captures the differences in transplanted kidney regions, and how this may differ among different patient traits such as age, sex, and primary diagnosis of kidney disease. It uses data from 821 unique kidney transplant recipients at a single institution over the span of five years across several ultrasound machines and settings. Using a fully supervised neural network, the transplanted kidney's regions are automatically extracted. In each resulting region of interest, seven statistical distributions are evaluated. In demonstrating a proof-of-concept and proposing a statistical distribution in the transplanted kidney, such information may be used in downstream tasks of segmentation, detection, or prediction. Furthermore, in developing an understanding of scatterer properties of transplanted kidneys non-invasively, we may yield insights into underlying disease states.

%%%%%%%%%%% MATERIALS AND METHODS
\section*{Materials and Methods}
\label{MaM}
Ultrasound data was retrospectively collected after Research Ethics Board approval (H19-02669). Patients included in this study were kidney transplant recipients receiving care at Vancouver General Hospital (Vancouver, BC) from 2014 to 2019. A total of 821 recipients were included. Ultrasound imaging performed by certified and trained sonographers followed standard imaging acquisition protocol which includes thorough Doppler investigation. B-mode images were recorded in this study. The scans were performed using curvilinear array transducers from  different machines and manufacturers: Philips, Siemens, GE, Acuson, SonoSite and Toshiba. In conjunction with the British Columbia Transplant Society, a province wide organization that coordinates care for organ transplant recipients, the anonymized patient demographic and characteristics data were obtained and linked to imaging data. 

As an overview, each ultrasound image is automatically segmented into three regions of interest (cortex, medulla, and central echogenic complex) using a trained neural network. For each region of interest, seven different distributions are fit. Statistical analysis is performed to identify distributions whose parameters are significantly different across each region. For these identified distributions, goodness of fit and the relative entropy is evaluated. Finally, the results are stratified by different patient characteristics and, after data processing, are assessed for significant correlations with the distribution parameters.

\begin{figure}
    \centering
    \includegraphics[width=\textwidth]{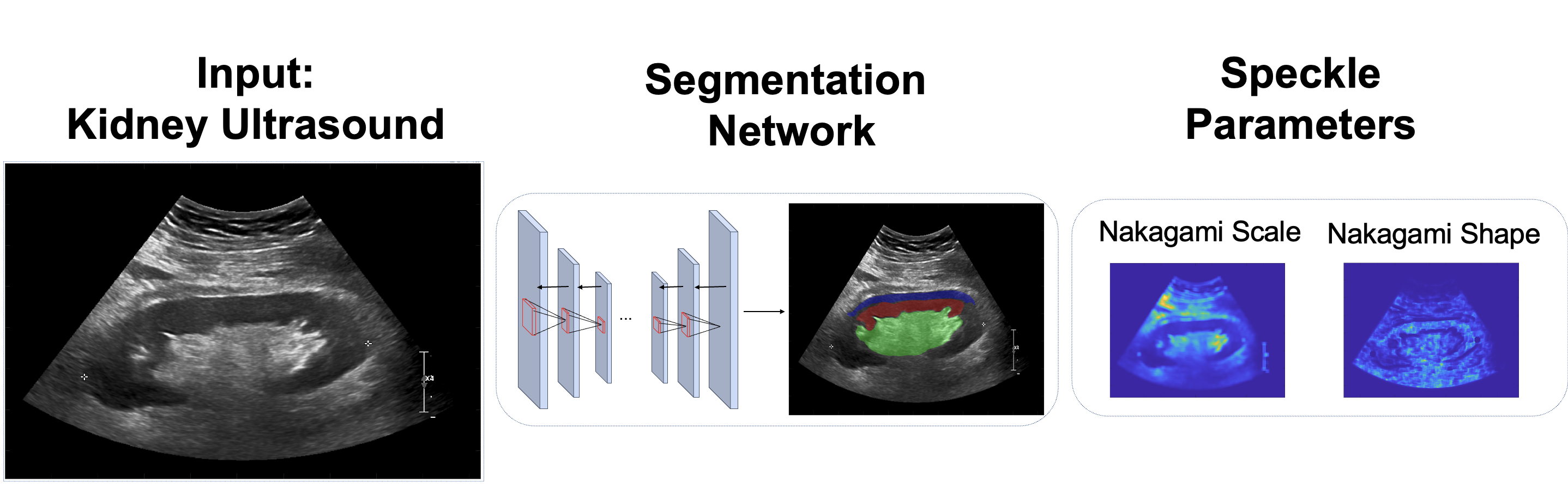}
    \caption{Illustrative overview of this paper. Ultrasound images are used as input into a segmentation network which generates multi-class masks. Within each mask, seven different probability distributions are calculated to generate a set of speckle parameters. The figure above uses the Nakagami parameters as an example.}
    \label{fig:flowchart}
\end{figure}

\subsection*{Probability Distributions Investigated}
One of the simplest and most frequently used probability distributions for speckle is the Rayleigh distribution.\citep{aysal2007rayleigh,sarti2005maximum,wagner1988fundamental} The Rayleigh distribution is a single-parameter model that assumes dense, random, and identical scatterers in tissue, allowing it to model fully developed speckle. \cite{shankar1996studies, goodman1975statistical} The continuous Rayleigh random variable $X$ has the following probability density function, where $\sigma$ is the scale factor.
\begin{equation}
    f(x,\sigma)=\frac{x}{\sigma} e^{(\frac{-x^2}{2\sigma^2})},
\end{equation}
In B-mode ultrasound imaging, $x$ is a pixel’s intensity on the ranging from 0 to 255. Here, $0 \leq \sigma \leq \infty$. However, as several authors have shown, there can exist non-Rayleigh distributed speckle depending on the scattering conditions.\citep{shankar1996studies,tuthill1988deviations} For example, the Rician distribution expands on the Rayleigh and allows for the introduction of periodically located scatterers.\citep{insana1986analysis} Closely related is the Nakagami distribution. This distribution is defined by two parameters the Nakagami (or shape) parameter m, and the scale factor, $\Omega$.\citep{cheng2001maximum} This distribution allows for more generality than the Rayleigh distribution while being more efficient to compute than the Rician distribution. A continuous Nakagami random variable $X$ has the probability density function
\begin{equation}
f(x,m,\Omega)= \frac{2m^m}{\Gamma(m) \Omega^m } x^(2m-1) e^(\frac{-m}{\Omega}x^2)
\end{equation}
where $\Gamma(*)$ is the Gamma function and $x$ is the pixel’s intensity. The Nakagami parameters are defined on the domains $\frac{1}{2} \leq m \leq \infty$ and $0 \leq \Omega \leq \infty$

Closely related to the Nakagami distribution is the Gamma distribution. This relation is defined such that if a random variable $Y \sim N(m, \Omega)$ then $Y^2 \sim Gamma(m, \frac{\Omega}{m})$. The Gamma distribution has shown excellent results in speckle modelling in empirical tests, such as in echocardiography, motivating its investigation in this work. \citep{nillesen2008modeling, tao2006evaluation}

In recent work by \citeauthor{parker2020burr}, the authors explored for new distributions that may better characterize the scattering properties of soft vascularized tissues.\citep{parker2020burr} These included the Burr, Pareto, and Lomax (Pareto Type II) distributions.\citep{parker2020burr} Notably, the Burr distribution has two different shape parameters $c$ and $k$, potentially allowing for broad range of potential probability density functions. The authors propose that in certain imaging conditions the fluid-filled vasculature within an organ may be the dominant scatterer in ultrasound, following a cylindrical shape rather than the conventionally assumed point-based or spherical shaped scatterers. The liver is used an exemplar organ in their work. Given that the kidney is a highly vascularized organ and smaller in size than the liver, the renal vasculature may have more significant influence in this organ, and may be a prominent scatterer in the transplanted kidney. 

In general, all of these distributions are left-skewed, heavy tailed, and have well-studied properties. The different distributions have demonstrated potential clinical utility in numerous applications. \citeauthor{tao2006evaluation} used the Gamma distribution to characterize speckle of the heart.\citep{tao2006evaluation} \citeauthor{hu2019acoustic} made use of the Nakagami distribution to characterize acoustic shadow regions caused by bones. \citeauthor{singla2022speckle} used speckle parametric maps as a type of data augmentation for training machine learning algorithms.\citep{singla2022speckle}. \citep{hu2019acoustic} \citeauthor{baek2021clusters} use the Burr distribution to distinguish between healthy and fatty livers, while \citeauthor{tsui2016acoustic} utilized the Nakagami distribution for fibrosis detection.\citep{baek2021clusters,tsui2016acoustic} To the best of our knowledge, these distributions have not been investigated in native or transplanted kidneys.

\subsection*{Region Extraction, Distribution Fitting and Statistical Analysis}      
For each patient’s B-mode scan (each containing ~200 frames), the frame obtained in a longitudinal view with the maximum cross-sectional area was extracted. This maximizes each of the three regions visible in the ultrasound frame. In this single frame, a neural network based on nnU-net was applied to extract three regional masks for each of the cortex, medulla, and central echogenic complex.

\begin{figure}
    \centering
    \includegraphics[width=\textwidth]{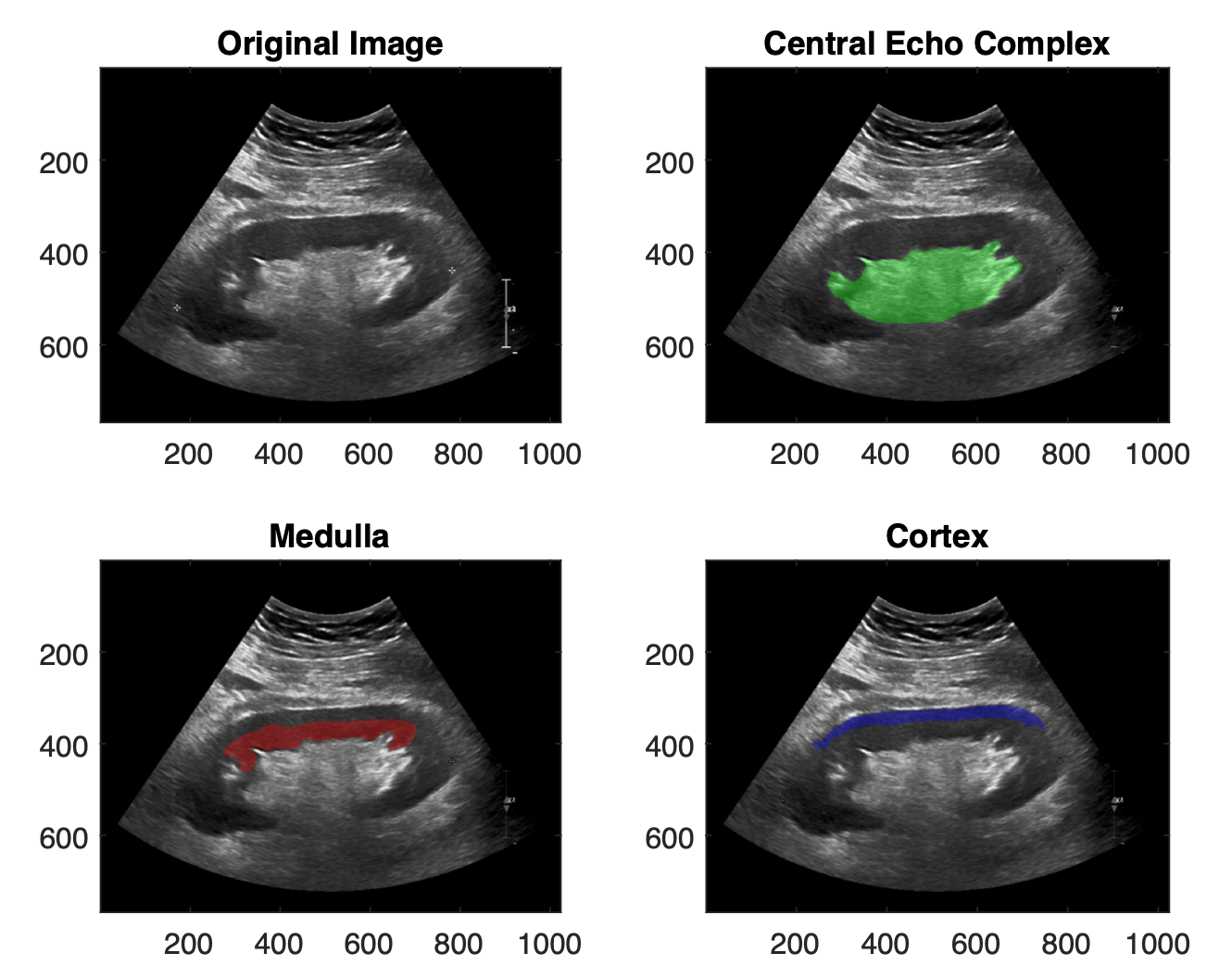}
    \caption{Example of each of the regional masks produced by the segmentation network.}
    \label{fig:classes}
\end{figure}

The details of the neural network’s implementation are reported by \citeauthor{isensee2021nnu}. \citep{isensee2019automated} The nnU-net is trained using a separate set of 514 kidney ultrasound images, using fine-grained polygon annotations from two expert sonographers each with over 20 years of expertise. The standard 80/20 training and testing ratio is adopted. No modifications to nnU-net’s data-adaptive methodology was used resulting in a standard two-dimensional U-net being trained with an Adam optimizer, combo loss of both Dice-Sorenson Coefficient and cross-entropy, an annealed learning rate of 0.01, and 500 epochs. The computations are carried out on a single NVIDIA Tesla V100 GPU. For each extracted regional mask, the seven distributions are each fit onto the intensity values. Where possible, efficient estimators for distribution parameters were used, otherwise the maximum likelihood estimator was used. For example, in the Nakagami distribution, the inverse normalized variance estimator was used. \citep{kolar2004estimator}. After grouping the estimated parameters in each region across all 821 patients, the parameters can be expected to follow a normal Gaussian distribution following the Central Limit Theorem. Normality was confirmed visually. Subsequently, for each distribution, Student’s t-test was used to evaluate significance difference in the parameters between regions in a pair-wise manner. An alpha value $\alpha$ of 0.05 was used for significance. However, given that a total of 42 comparisons on the same data were being performed, the likelihood of family-wise Type I error was high (88\%). Bonferroni correction was used to adjust for this, providing a corrected alpha value $\alpha_c$ of 0.0012.\citep{bland1995multiple}

\subsection*{Kullback-Leibler Divergences and Goodness of Fit}
After identifying the distributions with parameters that were significantly different between all three kidney regions, we evaluate the goodness of fit and relative entropy. To evaluate the quality of the estimated distributions to the underlying data, the goodness of fit is used. This is defined the sum of squares error across all images and masks for each of the distributions. To identify which of the distributions best separates the three regions, the Kullback-Leibler divergence (relative entropy) is computed. Given $x$ on the domain of $X$, the Kullback-Leibler divergence $D_{KL}$ is a measure of the difference from a query probability distribution $P(x)$ with respect to a reference probability distribution $Q(x)$. Maximizing the divergence of probability distributions between regions may yield better characterization, as it avoids potential redundancies. Conceptually, the Kullback-Leibler divergence represents new information gained from another distribution which increases if the distributions are different. As the Kullback-Leibler divergence is asymmetric, all pairs are compared. Only distributions with parameters that have a statistically significant difference between each region are used.
\begin{equation}
D_{KL}(P || Q)=  \sum_{x \in X} P(x) log(\frac{P(x)}{Q(x)})
\end{equation}
\subsection*{Stratification by Patient Characteristics}      
In addition to fitting different probability distributions to transplanted kidney regions, we sought to investigate how these distributions may differ based on transplant recipient characteristics. Based on the results of the Bonferroni correct t-test, Kullback-Leibler divergences, and goodness of fit metrics, we select one distribution for further investigation. For each image, we compared the distribution parameters in each of the cortex, medulla, and central echogenic complex with patient characteristics. We evaluated parameters for both recipient (age at transplantation, sex, body mass index, ethnicity, and primary diagnosis of kidney disease) as well as donor (living or deceased, and age). Any records with missing data were excluded from subsequent analysis. For the primary diagnosis causing kidney disease, there is a large variety of possible diseases. In a data processing step, disease instances with a frequency fewer than 10 were removed. This includes instances of amyloid, ANCA vasculitis, anti-GBM antibody disease (Goodpasture’s syndrome), benign nephrosclerosis, BK virus nephropathy, neurogenic bladder, congenital diseases, Fabry’s disease, granulomatosis with polyangiitis (Wegener’s disease), hepatorenal syndrome, Henoch-Schoenlein purpura, Meyer-Rokitansky syndrome, multi-system disease, obstructive uropathy, recessive polycystic kidney disease, thrombotic microangiopathy, and traumatic loss of kidney. For the categorical characteristics (sex, primary diagnosis, donor type, and ethnicity), we perform one-way ANOVA with a significance of $p \leq 0.05$ for assessing if a grouping differs significantly from another. In the event of a significant f-statistic for multiple groups, post-hoc tests are performed to identify the significantly different groups. For the numerical characteristics (age, and body mass index), two-tailed Pearson’s correlation coefficient was computed for each parameter-characteristic pair, using a significance of $p \leq 0.05$.

%%%%%%%%%%% Results
\section*{Results}
\label{Results}
Table 1 summarizes the parameters and if they were significantly different between regions. Only two distributions, Nakagami and Rayleigh, had parameters that achieved significant differences in all three regions. These two distributions were the only ones used for subsequent KL divergence and goodness of fit analysis. Table 2 summarizes the divergence values. 

\begin{figure}
    \centering
    \includegraphics[width=\textwidth]{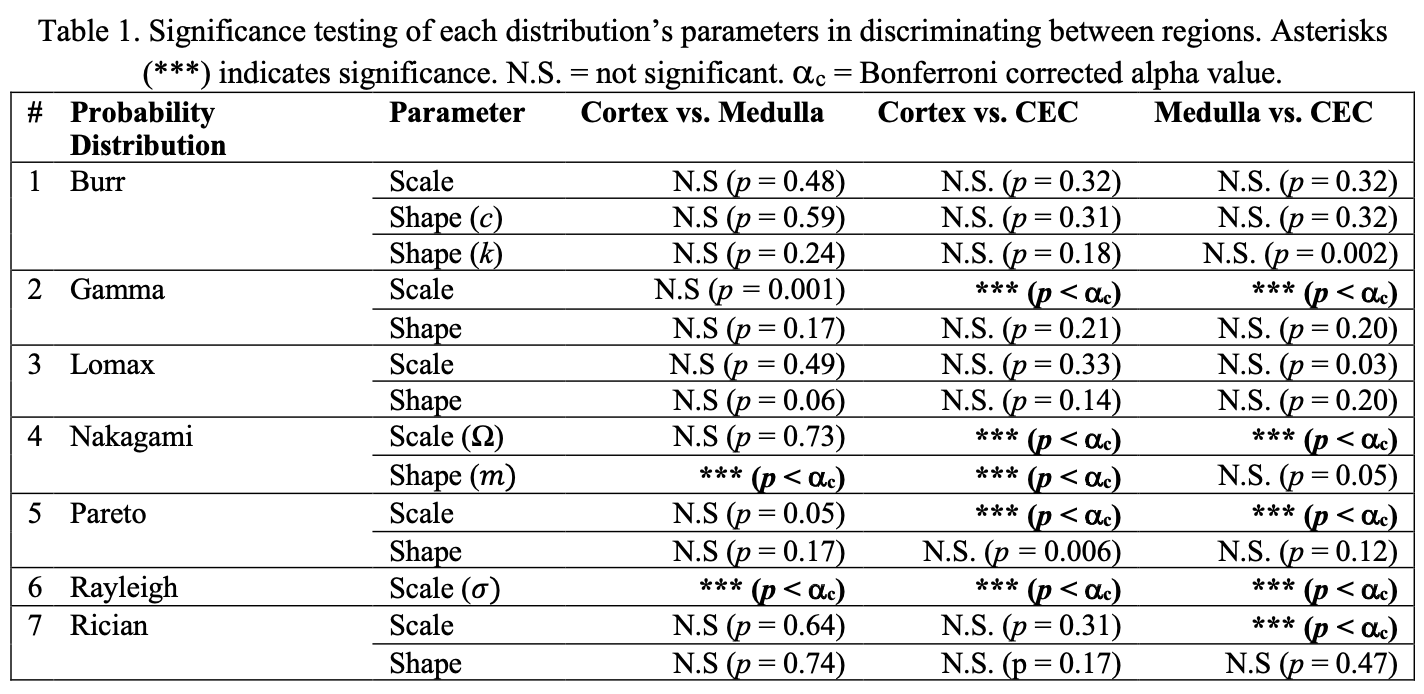}
    \caption*{}
    \label{fig:table1}
\end{figure}

\begin{figure}
    \centering
    \includegraphics[width=\textwidth]{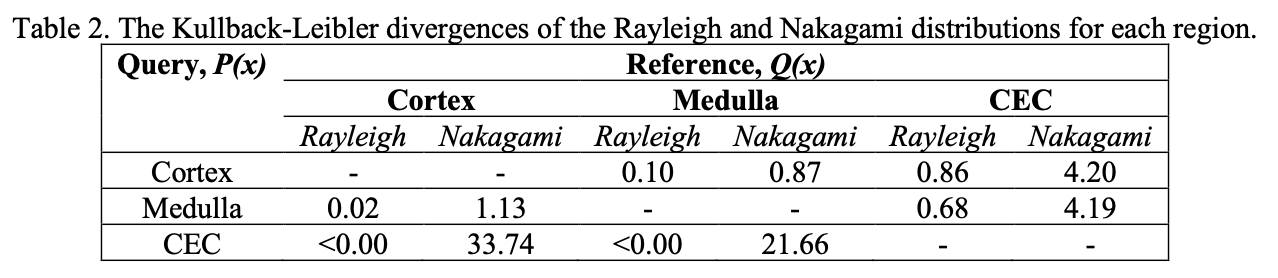}
    \caption*{}
    \label{fig:table2}
\end{figure}

Given the higher divergence values, the Nakagami distribution captures the distinctiveness of each region in a compelling manner. In comparing the cortex and medulla for example, two regions which are visually close together in echogenicity and difficult to segment, the Nakagami distribution has higher divergences compared to the Rayleigh distribution. Table 3 summarizes the goodness of fit for these distributions. 

\begin{figure}
    \centering
    \includegraphics[width=\textwidth]{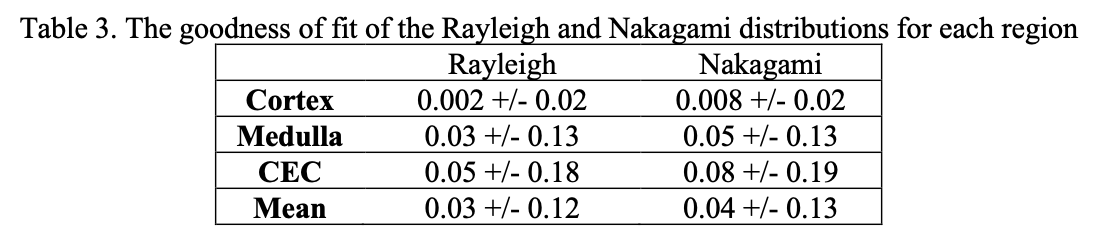}
    \caption*{}
    \label{fig:table3}
\end{figure}

We observe low fitting error in both distributions, with no significant differences in the sum of squares error for either one. While both Rayleigh and Nakagami distributions have excellent model fitting and divergence values, the Nakagami can be used in a more general manner than the Rayleigh. The Nakagami shape parameter, m, can be used to determined pre-Rayleigh, Rayleigh, and post-Rayleigh scattering conditions. We select this distribution for stratification.

When investigating subgroup populations of patients for differences, only recipient age and body mass index had significant correlations. The Nakagami shape and scale parameters in the central echogenic complex both demonstrated weak correlation with age (m: $\rho$ = -0.11, $p$ = 0.006;  $\Omega$: $\rho$ = 0.11, $p$ = 0.003). In the cortex, only the scale parameter demonstrated a weak correlation with age ($\Omega$: $\rho$ = 0.11, $p$ = 0.004) while the shape did not. In the medulla, no parameter correlated with age. Recipient body mass index had two significant weak correlations with the Nakagami distribution. The scale parameter in the central echogenic complex was weakly correlated with body mass index ($\Omega$: $\rho$ = 0.10, $p$ = 0.04) and the shape parameter in the medulla was also weakly correlated (m: $\rho$ = 0.08, $p$ = 0.04). All other subgroups, including recipient’s sex, ethnicity and primary diagnosis for end-stage kidney disease were not significantly different for any parameter in any region. Likewise, donor age and type were also not significantly correlated with any regional parameter.

%%%%%%%%%%% DISCUSSION
\section*{Discussion}
\label{Discuss}

In this study, we investigated seven different probability distributions for their ability to characterize three regions of the kidney in ultrasound images of adult kidney transplant recipients. Despite the nuanced relationships between several of the distributions, we found only two distributions (Rayleigh and Nakagami) had parameters that significantly differed between the kidney’s regions. Of these two, the Nakagami may be preferred due to its generality, efficiency, and fit.  The Nakagami shape parameter, $m$, can be used to define pre-Rayleigh, Rayleigh, and post-Rayleigh scattering conditions.\cite{shankar2000general}. This enables the Nakagami distribution to characterize a wider variety of conditions than the Rayleigh alone, which relies on fully developed speckle. The efficient estimators from \citep{kolar2004estimator} also make the parameters straightforward to compute. Given this however, it warrants to investigate further statistical distributions such as the generalized K or homodyned-K models. While they require intensive analytics to estimate the parameterization, the parameters are linked to signal coherency and may have specific physical meanings not captured by the Nakagami. It may also be possible to learn a distribution through machine learning techniques, rather than rely on symbolic modelling.

It is reassuring to see minimal impact to the Nakagami parameters when examining patient characteristics. While age and body mass index had significant correlations, the correlation was relatively weak. The reliance of fitting these distributions onto the B-mode, and hence enveloped data, makes them subject to variations in settings from different machines and transducers. For example, in the case of higher body mass index, a larger imaging depth and increased time-gain compensation may be needed to improve image quality. Alternatively, different post-processing steps and image enhancement algorithms from manufacturers may result in differing B-mode images. Ideally, the exploration of speckle distributions would be performed in radio-frequency data. Given our large sampling of B-mode images from multiple different machines (and settings), as well as the low fitting error, the Nakagami distribution remains an excellent model for speckle in the transplanted kidney.  

In terms of downstream application, the Nakagami parameters may assist as additional anatomical constraints in segmentation algorithms. By penalizing algorithms for inferring regions of interest for two different classes but have similar Nakagami parameters, the use of speckle modelling may help improve segmentations. Such parameters could be incorporated into loss functions as an additional term for example. \citeauthor{singla2022speckle} also demonstrated potential applications for the use of speckle parameters as an augmentation technique, another area where speckle modelling may be of use. \citep{singla2022speckle}

While the Nakagami parameters did not correlate with any individual primary disease, it is worth investigating what role these parameters have in screening for the presence of disease as well as predictive capabilities. Applicability to the native kidney also remains to be seen, but is warranted given the evident similarities in tissue structure between transplanted and native kidneys. Differences in organ location (the retroperitoneum versus the illiac fossa), host environment and immune response, and other factors may impact how well speckle is modelled. Finally, this investigation did not utilize the spatial distribution of speckle parameters nor included a temporal analysis. 

%%%%%%%%%%% Conclusions
\section*{Conclusions}
\label{Conclusions}
In an investigation of 821 kidney transplant recipients from a single institution, seven probability distributions were evaluated for their ability to characterize three kidney regions. The Nakagami distribution was found to be the most suitable for separating the regions in the transplanted kidney, with potential applications in characterizing the signal scatterers. We propose the Nakagami distribution be used to characterize transplanted kidneys regionally independent of disease etiology and most patient characteristics based on our findings. These findings may be useful in utilizing speckle characterization for non-invasive regional tissue characterization in the transplanted kidney. 

%%%%%%%%%%% ACKNOWLEDGEMENTS
\section*{Acknowledgements}
\label{Ack}
The authors acknowledge funding from the Natural Sciences and Engineering Research Council of Canada as well as the Kidney Foundation of Canada.

%%%%%%%%%%% REFERENCES
%% REFERENCE FORMATTING INSTRUCTIONS

%% All bibliography information should be included using a 'thebibliography' environment.  Most authors will find it easiest to create a .bbl file using the commands \bibliographystyle{} and \bibliography{} and then copy and paste the contents of the .bbl file into the .tex file below, but before the figure captions section.  Examples for using the \bibliographystyle and \bibliography commands are listed below.  

%% Do not remove the page break here.
\pagebreak

%% References with bibTeX database, use this to create a .bbl file
\bibliographystyle{UMB-elsarticle-harvbib}
\bibliography{bib}

%%%%%%%%%%% FIGURE CAPTIONS

%% Include only the figure captions here (not the figures).  Figures are uploaded separately in the online Elsevier Editorial Submission process.

%% Do not remove the page break here.
\pagebreak

\section*{Figure Captions}

\begin{description}
\item[Figure 1:] Overview of key methods. For an input ultrasound image, the three regions of the kidney are extracted via a neural network. Seven different probability distributions are fitted onto each region, and statistical analysis is performed after fitting. The Nakagami distribution is one of the seven. Subsequent subgroup analysis is performed on patient factors to evaluate any differences.

\item[Figure 2:] The three regions of the kidney that are extracted by the nnU-net segmentation. (Top left) the anonymonized kidney ultrasound image, (top right) the central echogenic complex which constitutes several anatomical structures but visually in ultrasound appears amalgamated, (bottom right) the cortex, a thin layer on the outermost portion of the kidney, and (bottom left) the medulla, the innermost portion of the renal parenchyma.

\end{description}

%%%%%%%%%%% TABLES AND TABLE CAPTIONS

%% Since the tabular format is often difficult to work with for complex tables, authors may also choose to create their tables with another program.  Each table and it's corresponding caption should then be saved as a pdf.  Each pdf should then be uploaded separately during the online submission process.  If doing so, all of the text below concerning tables and table captions should be commented out.

%% Do not remove the page break here.
\pagebreak

\section*{Tables}

\end{document}